\documentclass[longbibliography,floatfix,12pt,a4paper]{iopart}

\usepackage[nospace]{cite}

  \expandafter\let\csname equation*\endcsname\relax
  \expandafter\let\csname endequation*\endcsname\relax 
    
\usepackage{bm,amsmath,amssymb,graphicx,subfigure}

\usepackage[usenames,dvipsnames]{xcolor}

\definecolor{darkgray}{gray}{0.4}

\usepackage[abs]{overpic}
\newcommand{\uvc}[1]{\bm{\mathrm{\hat #1}}} 

\newcommand{\bX}{{\bf X}}
\newcommand{\bJ}{{\bf J}}

\usepackage{ulem}
\begin{document}

\title[Whirling skirts]{Whirling skirts and rotating cones}

\author{Jemal Guven}
\eads{\mailto{jemal@nucleares.unam.mx}}
\address{Instituto de Ciencias Nucleares, Universidad Nacional Aut\'{o}noma de M\'{e}xico;\\
Apdo. Postal 70-543, 04510 M\'{e}xico D.F., M\'{e}xico}
\author{J A Hanna}
\eads{\mailto{hannaj@vt.edu}}
\address{Department of Engineering Science and Mechanics, Virginia Polytechnic Institute and State University;\\
Norris Hall (MC 0219), 495 Old Turner Street, Blacksburg, VA 24061, U.S.A.}
\author{Martin Michael M\"{u}ller}
\eads{\mailto{martin-michael.mueller@univ-lorraine.fr}}
\address{Equipe BioPhysStat, ICPMB-FR CNRS 2843, Universit\'{e} de Lorraine;\\
1, boulevard Arago, 57070 Metz,  France}
\address{CNRS, Institut Charles Sadron;\\
23, rue du Loess BP 84047, 67034 Strasbourg, France}

\date{\today}

\begin{abstract} 
Steady, dihedrally symmetric patterns with sharp peaks may be observed on a spinning skirt, lagging behind the material flow of the fabric.  These qualitative features are captured with a minimal model of traveling waves on an inextensible, flexible, generalized-conical sheet rotating about a fixed axis.  Conservation laws are used to reduce the dynamics to a quadrature describing a particle in a three-parameter family of potentials. One parameter is associated with the stress in the sheet, aNoether is the current associated with rotational invariance, and the third is a Rossby number which indicates the relative strength of Coriolis forces.  Solutions are quantized by enforcing a topology appropriate to a skirt and a particular choice of dihedral symmetry.  A perturbative analysis of nearly axisymmetric cones shows that Coriolis effects are essential in establishing skirt-like solutions.  Fully non-linear solutions with three-fold symmetry are presented which bear a suggestive resemblance to the observed patterns.
\end{abstract}


Capturing the motion of a flexible sheet or solid membrane, be it only spinning about a fixed axis as embodied in a whirling skirt of fabric, presents challenges even in simulations \cite{Goldenthal07}. The dynamics of its folds are typically observed as transient events.  Yet the hypnotic ritual of the ``whirling dervish'', credited to the 13th century Persian poet and mystic 
 Rumi, provides us with observations of emergent steady state patterns in this system.
 A survey of videos found on the internet (for example \cite{UrusoffRamosSkirtsYouTube}) reveals unmistakable, if imperfect, stable configurations of skirts in a rotating frame.  These appear as dihedral waveforms with usually three, but occasionally four or more, lobes consisting of gentle troughs between remarkably ``sharp'' peaks--- smooth but with a large zonal curvature.  On close examination, it is apparent that these shapes lag behind the fabric; the material has a zonal surface velocity.   The still photograph in Figure \ref{coordinates} contains a few waves, although it cannot demonstrate the zonal flow of material.  The skirt itself is neither pleated nor appreciably rigid with respect to bending, so that when not rotating it resembles a hanging drape or tablecloth. 

Guided by these observations, we examine a simple model of a rotating body, balancing inertia and the constraint on the two-dimensional metric implied by an inextensible fabric.  We deliberately neglect several physical realities, among them the stretching of the material, its bending stiffness, its interaction with the air, gravity, and the effect of an uneven mass density provided by the weight of additional fabric along the hem.  Material at the edge of a one-meter-radius skirt revolving at one hertz experiences accelerations of about four times Earth gravity, so we ignore the latter at our quantitative peril.  This risk is offset, however, by the gain in mathematical and qualitative simplicity.  The point is not to make a better dancer or to provide a detailed description of the physics, but to understand the backbone of the dynamical system.  With such a simple model--- it is indeed the simplest possible, having only two terms in the equation of motion--- we can explore curvilinear geometries for which there is a direct coupling between in-surface stresses and dynamics.  Thus, we probe the importance of geometric nonlinearities in sustaining the observed patterns.  In this context, it should be remarked that the emergence of ``standing waves'' on stiff plates such as sawblades, turbines and hard disks, spinning in vacuum or subject to aeroelastic interactions, was long ago recognized as a harbinger of failure \cite{LambSouthwell21, Campbell24, Advani67, Renshaw94}.  This behavior has been studied either from the standpoint of small amplitude vibrations about a planar configuration, or by artificially imposing harmonic \emph{ans\"{a}tze}.  Our treatment of more flexible sheets, featuring a generalized-conical family of skirts,  exhibits a richer shape space that includes features  quite suggestive of the observed wave patterns.  Much of this space is inaccessible if attention is confined to perturbations of a planar disk.  

Our equations of motion can be viewed as the extension for higher dimensional objects of the equations for a classical string \cite{Routh55} describing a one-dimensional inextensible body moving in two or more dimensions.  Yet despite the fundamental nature of such equations, we are unaware of their prior appearance in the literature, a fact worthy of remark in its own right.  After all, they are presumably some thin-body limit of those describing the elastodynamics of an object with the same dimensions as its embedding space.  It is perhaps more helpful to think of the motion as metrically constrained, \emph{i.e.} rigid body, Newtonian dynamics where the rigid metric is of lower dimension than the embedding space inducing it.  Within this vast range of possible dynamics, one can ask if there exist rigid or otherwise shape-preserving motions of such extended, flexible bodies.  The solutions presented in this study are of this type.

Our treatment combines two complementary approaches.  As byproducts of elementary calculations, we find the stresses in the sheet, which serve as Lagrange multipliers for its metric.  More abstractly and indirectly, the symmetry of the action with respect to rotations about an axis is used to reduce the dynamics to a quadrature describing the motion of a fictitious particle in a three-parameter potential.  Combining the two viewpoints facilitates the interpretation of these parameters.

We first examine proper rigid rotations of a generalized-conical sheet. This  leads to a two-parameter family of shapes whose description lends itself to a simple exposition. However, such motions are too restrictive to describe whirling skirts.
We next exploit the rotational Killing field of our axisymmetric metric to 
allow the sheet material to flow within the moving surface. This introduces a third parameter, generates a wider variety of shapes through Coriolis effects, and provides some wiggle room to satisfy highly restrictive implicit global constraints on the topology and deficit angle of the cone.  In particular, dervish-like solutions become possible.  By numerically integrating the quadrature, we provide a few examples, reserving a detailed analysis of the solution space for a second paper.  A perturbative analysis of nearly axisymmetric cones is presented in \ref{smallosc}.


\section{Basic formulation}

The equations of motion for an inextensible, perfectly flexible sheet with uniform mass density $\mu$ are given by
\begin{align}\label{motion}
	\mu\partial_t^2\bX &= \nabla_\alpha \left( \sigma^{\alpha\beta} \partial_\beta \bX \right) \, , \\
	\partial_\alpha\bX \cdot \partial_\beta\bX &= a_{\alpha\beta} \, , 
\end{align}
where $\alpha, \beta \in \{1, 2\}$, and $\bX\left( \xi^\alpha, t \right)$ is a Cartesian position vector describing the location of the sheet as a function of time $t$ and a suitable pair of local surface coordinates $\xi^\alpha$. The stress tensor $\sigma^{\alpha\beta}$ is a local Lagrange multiplier enforcing the time-independence of a two-dimensional metric $a_{\alpha\beta}$, and $\nabla$ is a covariant derivative constructed with this metric. 
The specific geometry we will consider is that of a generalized cone (Figure \ref{coordinates}),
parametrized by the distance $r$ from the conical singularity and the distance $s$ along the curve traced by the intersection of the cone with a sphere of unit radius centered on the singularity.  Hence, the metric of the surface is given by: 
\begin{equation}\label{metric}
	a_{\alpha\beta}(r, s) = \left(\begin{array}{cc}
				1 & 0\\
				0 & r^2\\
				\end{array}\right) \, ,
\end{equation}
flat everywhere except at $r=0$.  The strength of the singularity does not explicitly appear in \eqref{metric} because $s$ is a dimensionless zonal (azimuthal) coordinate, akin to an angle, while the meridional (radial) coordinate $r$ has units of length.


\section{Rigid skirts}\label{rigidskirt}

\begin{figure}[here]

\subfigure{
	\begin{overpic}[width=2.5in]{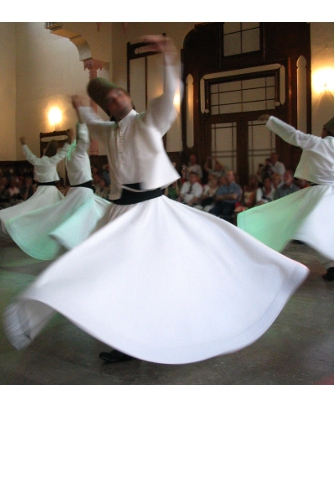}
	\end{overpic}
	}
\hspace{-.3in}
\raisebox{.45in}{
\subfigure{
	\begin{overpic}[width=4in]{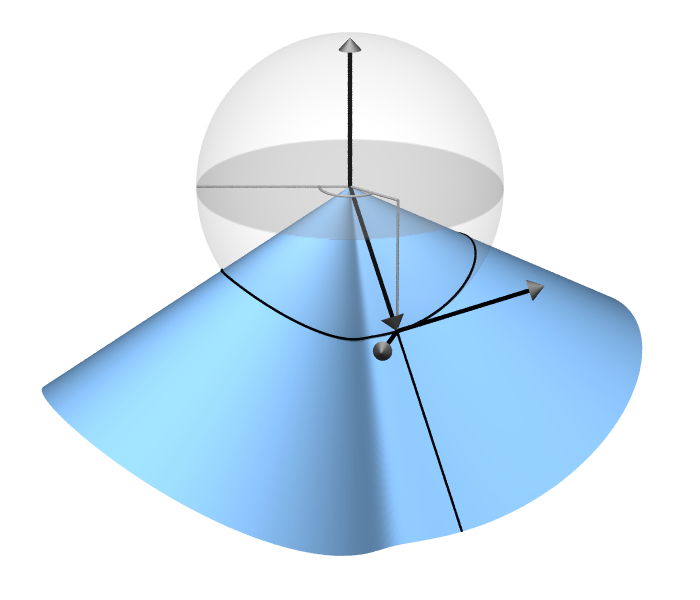}
	\put(167,135){\color{darkgray}$Z$}
	\put(155,170){\color{darkgray}$R$}
	\put(135,155){\color{darkgray}$\Phi$}
	\put(120,105){$s$}
	\put(178,75){$r$}
	\put(148,210){$\uvc{z}$}
	\put(148,115){$\uvc{u}$}
	\put(148,90){$\uvc{n}$}
	\put(210,112){$\uvc{t}$}
	\end{overpic}
	}
	}\\
\vspace{-.8in}
\caption{Left: Waves on dervishes' skirts.  Photo adapted from the work of Matt Lingard (\texttt{www.flickr.com/photos/madrattling/2395433079/}), modified to blur faces.  Right: An illustration of the coordinates and orthonormal triad used in this paper.  These are drawn on a generalized cone, a shape specified by a curve on the unit sphere. }
\label{coordinates}
\end{figure}

We first seek steady solutions for a rigid cone rotating with frequency $\omega$ about a fixed axis $\uvc{z}$, using the cylindrical representation:
\begin{align}
	\bX &= r\uvc{u}(s, t) \, , \label{cyl1} \\
	\uvc{u}(s, t) &= 
	\left(\begin{array}{c}
	R(s)\cos[\omega t + \Phi(s)]\\
	R(s)\sin[\omega t + \Phi(s)]\\
	Z(s)
	\end{array}\right) \, . \label{cyl2}
\end{align}
We will also use an orthonormal triad comprising $\uvc{u}$, $\uvc{t} = \partial_s\uvc{u}$, and $\uvc{n} = \uvc{u} \times \uvc{t}$, whose evolution along the zonal coordinate $s$ is given by
\begin{equation}
	\partial_s \left(\begin{array}{c}
			\uvc{u}\\
			\uvc{t}\\
			\uvc{n}
			\end{array}\right) = \left(\begin{array}{ccc}
							0 & 1 & 0\\
							 -1 & 0 & -k \\
							0 & k & 0
							\end{array}\right)
	\left(\begin{array}{c}
	\uvc{u}\\
	\uvc{t}\\
	\uvc{n}
	\end{array}\right) \, .
\end{equation}
The vector $\uvc{u}$ traces out a curve of dimensionless geodesic curvature $k(s)$ on the unit sphere, $\uvc{t}$ is this curve's unit tangent, and $\uvc{n}$ is the unit normal to the cone \cite{GuvenMuller08}.  The constraints on the lengths of $\uvc{u}$ and $\uvc{t}$ require
\begin{align}
	R^2 + Z^2 &= 1 \, , \label{firstconstraint}\\
	(\partial_sR)^2 + (\partial_sZ)^2 + (R\partial_s\Phi)^2 &= 1 \, . \label{secondconstraint}
\end{align}
Thus, in this representation, $k$ may be written in terms of any of the functions $Z$, $R$, or $\partial_s\Phi$, and the conical immersion $\bX$ may be fully specified by either $R$ or $Z$.  Figure \ref{coordinates} illustrates the coordinate systems and the $(\uvc{u}, \uvc{t}, \uvc{n})$ triad.

If the boundaries are cut appropriately along cone coordinate lines, we may assume that the stress is diagonal in these coordinates, $\sigma^{rs}=\sigma^{sr}=0$.  The resulting equation of motion, obtained from \eqref{motion}, 
\begin{equation}\label{motion2}
	-\mu\omega^2 r^2R 	\left(\begin{array}{c}
					\cos[\omega t + \Phi]\\
					\sin[\omega t + \Phi]\\
					0
					\end{array}\right) = \partial_r \left( r\sigma^{rr} \uvc{u}\right) + \partial_s \left( r^2 \sigma^{ss} \uvc{t} \right) \, ,
\end{equation}
is tractable.  Note that any axisymmetric cone aligned with the $\uvc{z}$ axis so as to have constant $R$ and $Z$ is a solution: $\Phi = \frac{s}{R}$, $k = -\frac{Z}{R}$, $\sigma^{ss} = \mu\omega^2R^2$, $\partial_r(r\sigma^{rr})=0$.  

Projection of the equation of motion on $\uvc{t}$ may be integrated in $s$ to obtain
\begin{equation}\label{tproj}
	2\sigma^{ss} = -\mu\omega^2R^2 + C(r) \, ,
\end{equation}
while projection on $\uvc{u}$ and integration in r leads to
\begin{equation}\label{uproj}
	2r\sigma^{rr} = D(s) + 2 \int \left(\sigma^{ss} - \mu\omega^2R^2 \right)r^2 dr \, ,
\end{equation}
where $C(r)$ and $D(s)$ are arbitrary functions of $r$ and $s$, respectively.  Examination of the $r$ dependence of the projection on $\uvc{n}$ indicates that $C(r)$ must be a constant $C$ and, as a result,  the integral in \eqref{uproj} can be evaluated explicitly.  With boundaries at constant values of $r$, we may tune $D(s)$ to make the outer boundary free of forces, and thus determine the forces induced on the inner boundary.
The $\uvc{n}$ projection may also be manipulated into a second order ODE for $Z$ or $R$ or $\partial_s\Phi$.  Yet there is a simpler route to a simpler ODE.  Projection on $\partial_t\uvc{u}$ or the velocity vector $\partial_t \bX$, followed by integration over $s$ defines a constant
\begin{equation}\label{Jz}
	2\sigma^{ss}R^2\partial_s\Phi = 2\sigma^{ss}\uvc{n}\cdot\uvc{z} = \tfrac{2}{\omega}\sigma^{ss}\uvc{t}\cdot\partial_t\uvc{u} \equiv J_z \, .
\end{equation}
After appropriate substitutions for $R$ and $\Phi$ using (\ref{firstconstraint}-\ref{secondconstraint}) and the definitions $\hat{J}_z \equiv \frac{J_z}{\mu\omega^2}$ and $\hat{C} \equiv \frac{C}{\mu\omega^2}$, the relations in \eqref{Jz} reduce to a first order ODE for $Z$,
\begin{equation}\label{zode}
	\left(\partial_s Z \right)^2 = 1 - Z^2 - \left( \frac{\hat{J}_z}{1 - Z^2 - \hat{C}} \right)^2 \, .
\end{equation}
A change of variables to $Z^2$ or $R^2$ will result in a similar first order ODE with a quartic on the right hand side, implying that implicit solutions may be written in the form of an elliptic integral.  Equation \eqref{zode} describes the orbit of a particle with position $Z$ in a potential that depends qualitatively on the sign of $\hat{C}-1$.
  
The constants of integration $J_z$ and $C$ are closely related to two constants that appeared in an earlier analysis of rotating strings by one of the authors  \cite{Hanna13}, where they were named $c_2$ and $c_3$.  The value of $C$ is clearly connected to the stresses in the sheet through the relations \eqref{tproj} and \eqref{uproj}.  Additionally, $J_z$ may be identified with the Noether current corresponding to rotational symmetry of an action or Lagrangian from whence we could have derived our equations of motion \eqref{motion} or \eqref{motion2}.  That is, consider the functional
\begin{equation}
	L = \int_\mathcal{S} \sqrt{a}\left[ \mu \omega^2\left(\uvc{z}\times\bX\right)\cdot\left(\uvc{z}\times\bX\right) - \sigma^{\alpha\beta}\left(\partial_\alpha\bX\cdot\partial_\beta\bX - a_{\alpha\beta}\right)\right] \, ,
\end{equation}
corresponding to rotation about $\uvc{z}$ of an object with fixed metric $a_{\alpha\beta}$.  Under a small variation $\delta \bX$ we have
\begin{equation}\label{variation}
	\delta L = 2\int_\mathcal{S} \sqrt{a}\left[ \mu \omega^2 \left(\bX - \bX\cdot\uvc{z}\uvc{z} \right) \cdot \delta \bX + \nabla_\alpha \left(\sigma^{\alpha\beta}\partial_\beta\bX \right) \cdot \delta \bX - \nabla_\alpha\left(\sigma^{\alpha\beta}\partial_\beta\bX \cdot \delta \bX \right) \right] .
\end{equation}
For a rigid rotation about $\uvc{z}$, $\delta \bX \propto \uvc{z} \times \bX$, of a cone with diagonal stress, the conserved (boundary) term is
\begin{equation}
	 -2\sqrt{a}\left[\sigma^{\alpha\beta}\partial_\beta\bX\cdot\left(\uvc{z} \times \bX\right)\right] 
	=  -2r\left[ \; \uvc{z} \cdot \left( \bX \times \sigma^{ss}\partial_s\bX \right)\right] 
	=  -r^3 J_z\,.
\end{equation}
The constant  is named $J_z$  to indicate that it is the projection onto the rotational axis $\uvc{z}$ of the conserved vector $\bJ$ associated with the full rotational invariance of static cones \cite{GuvenMuller08}. 

\begin{figure}[here]
\subfigure{
	\begin{overpic}[width=3in]{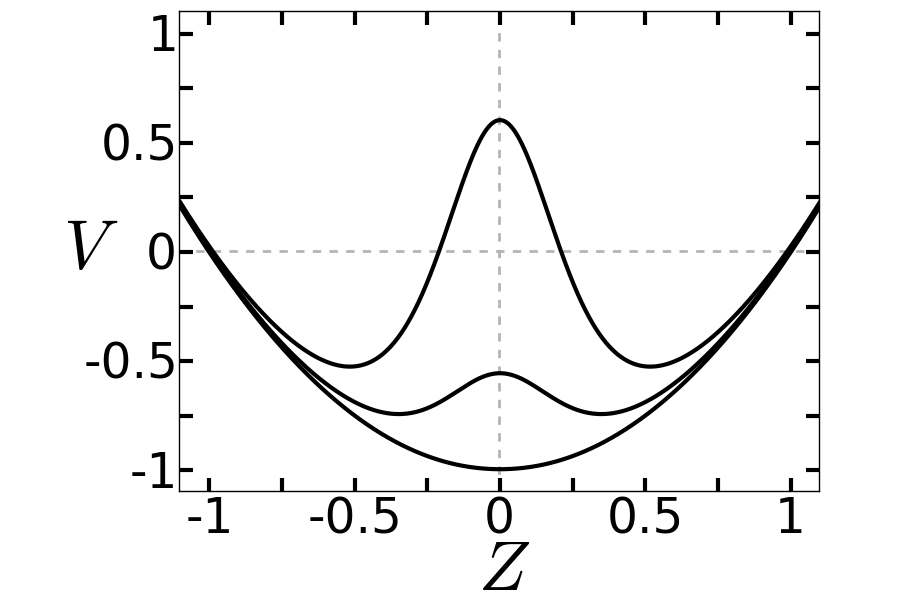}\label{first}
	\put(45,120){\Large{\subref{first}}}
	\end{overpic}
	}\hspace{-.25in}
\subfigure{
	\begin{overpic}[width=3in]{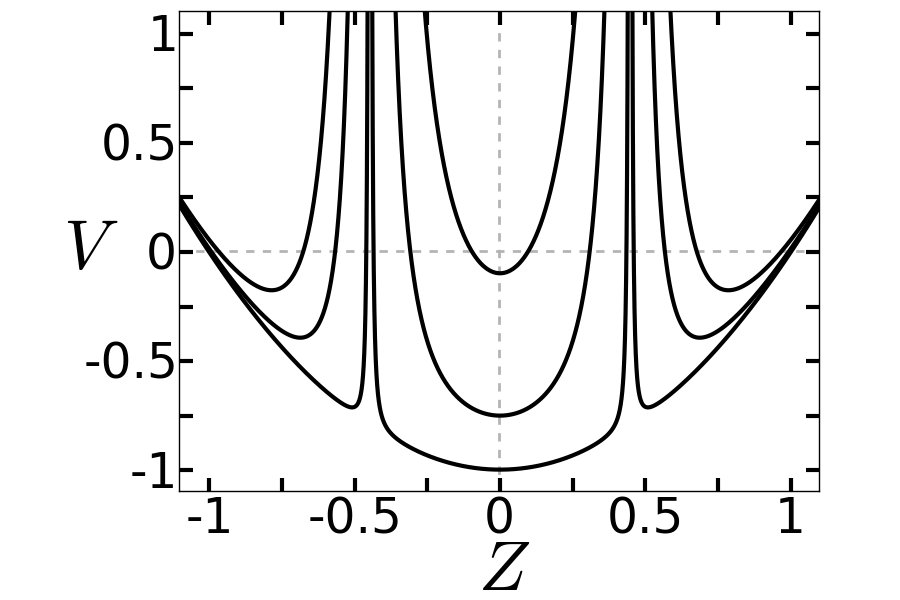}\label{third}
	\put(45,120){\Large{\subref{third}}}
	\end{overpic}
	}\\
\subfigure{
	\begin{overpic}[width=3in]{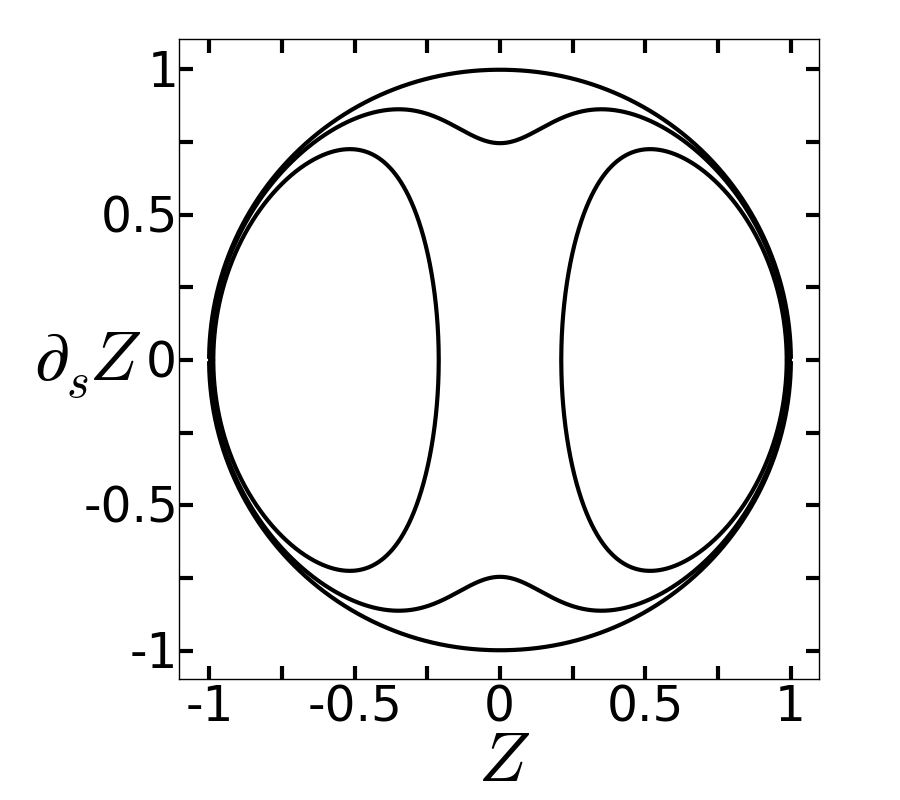}\label{second}
	\put(45,160){\Large{\subref{second}}}
	\end{overpic}
	}\hspace{-.25in}
\subfigure{
	\begin{overpic}[width=3in]{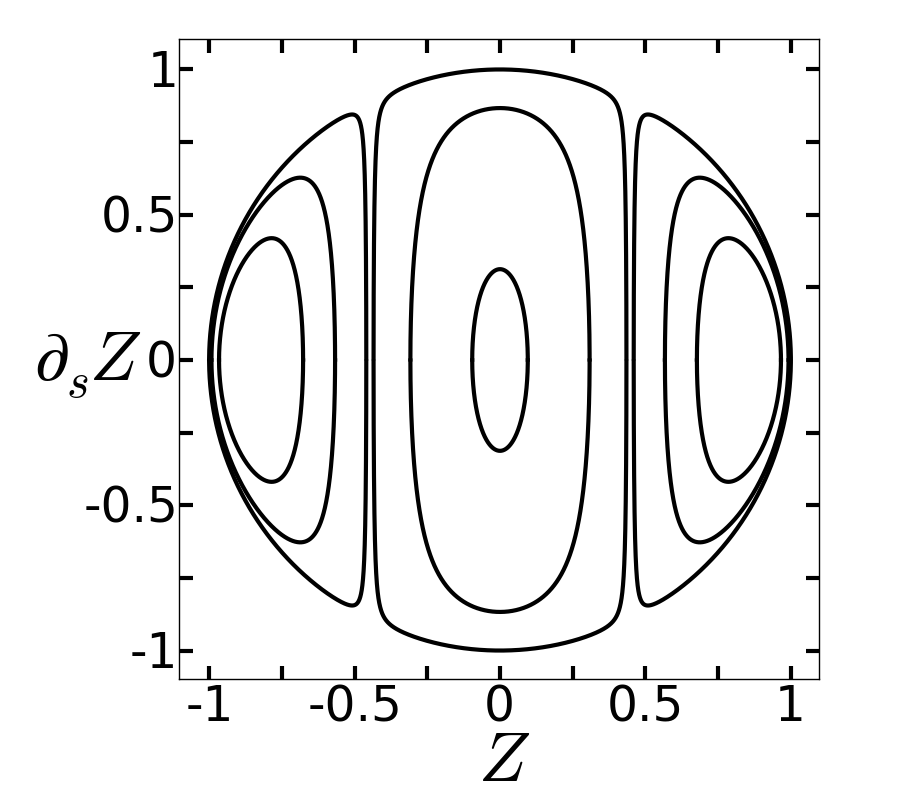}\label{fourth}
	\put(45,160){\Large{\subref{fourth}}}
	\end{overpic}
	}
\caption{Potentials $V$ \subref{first}, \subref{third} and phase portraits \subref{second}, \subref{fourth} for motion of the ``particle'' $Z$ according to equation \eqref{zode}, with definition of $V$ as in the text.  All orbits lie inside the circle $Z^2 + (\partial_sZ)^2=1$ and intersect the $Z$ axis at the real roots of $(1-Z^2)(1-Z^2-\hat{C})^2 = \hat{J}_z^2$.  Increasing $\hat{J}_z$ increases the potential and shrinks the orbits.  \subref{first} and \subref{second} Single and double well cases $\hat{C}=1.15$, $\hat{J_z} = (0.01, 0.1, 0.19)$,  \subref{third} and \subref{fourth} Triple well cases $\hat{C}=0.8$, $\hat{J_z} = (0.01, 0.1, 0.19)$, with barriers at $Z = \pm \sqrt{1-\hat{C}}$ .}
\label{phase}
\end{figure}

Let us return to the potential, which for simplicity of exposition we define as the negative of the right hand side of \eqref{zode}, as per $\left(\partial_sZ\right)^2 = -V$ with the usual factor of two absorbed into our definition of $V$. For $(\hat{C} -1)^3 \ge 2\hat{J}_z^2$, the potential $V$ has a single well, while for smaller $\hat{C}$ still greater than unity, it has a smooth double well.  When the barrier height between the wells becomes positive, it is insurmountable and corresponds to a separatrix in phase space.  Similarly, a well becomes inaccessible if its minimum becomes positive.  When $\hat{C} < 1$, there is a qualitative change to three wells separated by infinitely high barriers at $Z = \pm \sqrt{1-\hat{C}}$.  The zonal stress $\sigma^{ss}$ is compressive in the central well and tensile in the outer wells, as may be seen from inspection of \eqref{tproj}.  Because $Z \le 1$ by definition, the outer wells cease to exist when $\hat{C} =0$.  Potentials and phase portraits corresponding to \eqref{zode} are shown in Figure \ref{phase} for several values of the parameters $\hat{C}$ and $\hat{J}_z$.  All orbits are contained within the circle $Z^2 + (\partial_sZ)^2=1$, which is itself a limiting solution when $\hat{C} > 1$ and $\hat{J}_z \rightarrow 0$, corresponding to a flat sheet rotating about a line contained within itself.  

For the surfaces we describe to be disk- or rather skirt-like, complete orbits of the dynamical system \eqref{zode} must be commensurate with the corresponding winding of the surface generators about the $\uvc{z}$ axis of the embedding space, without self-intersections.  Fixing one parameter, either $\hat{C}$ or $\hat{J}_z$ or the cone angle representing the length of material along a curve at a fixed meridional distance, this extrinsic closure will act as a quantization condition on the other parameter.  The self-intersection issue is more difficult to treat.  It is not a concern for the two-parameter rigid skirt, but we will need to revisit it when we examine the full three-parameter flowing skirt.

As relationships between the parameters and variables are implicit, closure is usually implemented  numerically.  Inversion of equation \eqref{zode} tells us the range of $s$ that will take $Z$ from its minimum to maximum values or \emph{vice versa}, as given by a co-orbiting pair of roots of 
\begin{equation}\label{cubic}
	(1-Z^2)(1-Z^2-\hat{C})^2 = \hat{J}_z^2 \, .
\end{equation}
For an $n$-lobed skirt,
\begin{equation}\label{phi}
	\Phi = \int^{s} \!\!\! d\tilde{s} \, \frac{\hat{J}_z}{[1-Z^2(\tilde{s})] [1-Z^2(\tilde{s})-\hat{C}]} = \frac{\pi}{n} \, 
\end{equation}
for a single such half-traversal of an orbit.  The total $s$ over all the lobes of the skirt must be the cone angle $2\pi + \Delta$, where $\Delta$ is the excess or deficit angle whose magnitude is the strength of the metric singularity.  These totals for $s$ and $\Phi$ constitute two integral constraints on the solutions.  The up-down symmetric solutions of a central well must have an excess angle ($\Delta > 0$), but the converse need not be true. Expression \eqref{phi} shows that $\Phi$ is monotonic in $s$, so any self-intersection other than periodic closure will not occur for integer-valued $n$.  

An explicit analytical description is possible for configurations that correspond to small perturbations around axisymmetric shapes. 
In this approach, multi-lobed rigid skirt solutions only occur as perturbations about a planar sheet, shapes whose zonal evolution oscillates symmetrically about the equator.  The interesting shapes of the dancer's skirt are clearly up-down asymmetric and lie entirely in one hemisphere; their description requires the addition of a third parameter.  Perturbative solutions are described in detail in \ref{smallosc}.  

In the following section, we recall our observation of material motion distinct from that of the steady pattern, and consider the corresponding additional  degree of freedom.


\section{Flowing skirts}\label{flowingskirt}

\begin{figure}[here]
  \begin{overpic}[width=3.75cm]{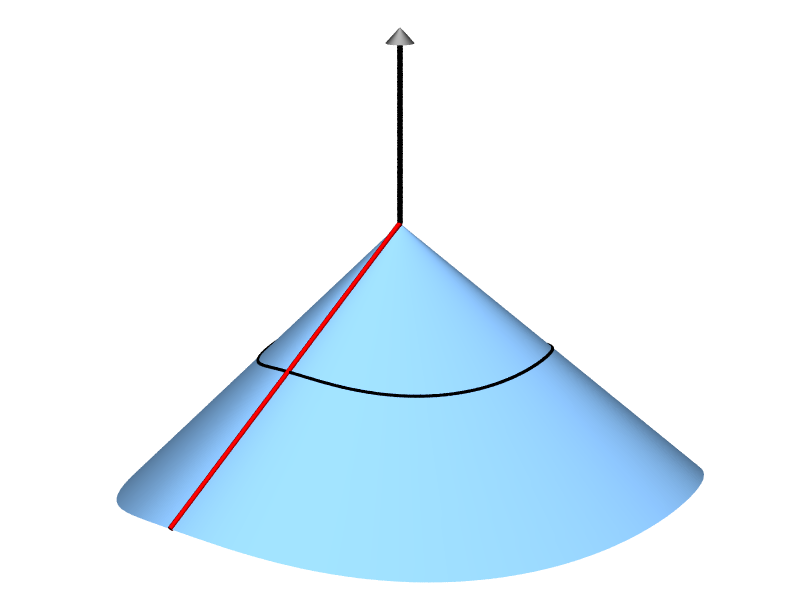}
	\put(100,40){$\Rightarrow$}
  \end{overpic}
  \begin{overpic}[width=3.75cm]{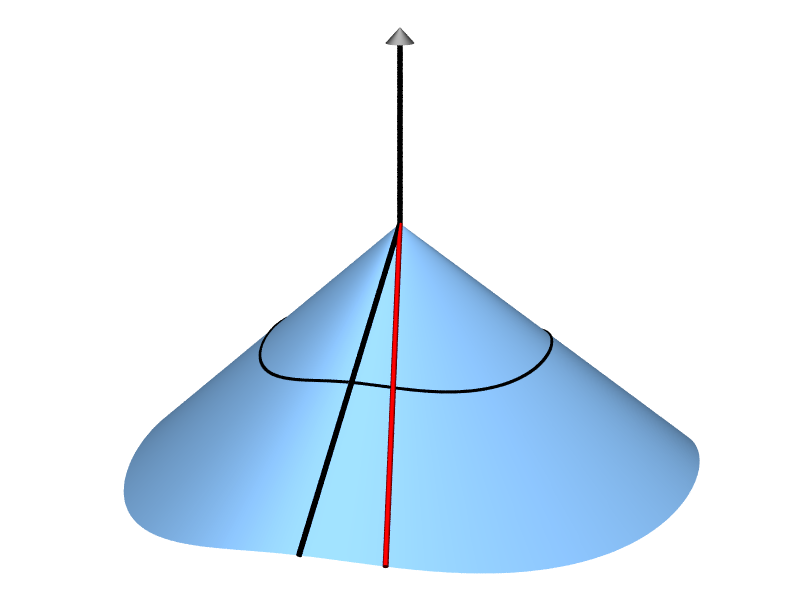}
	\put(100,40){$\Rightarrow$}
  \end{overpic}
  \begin{overpic}[width=3.75cm]{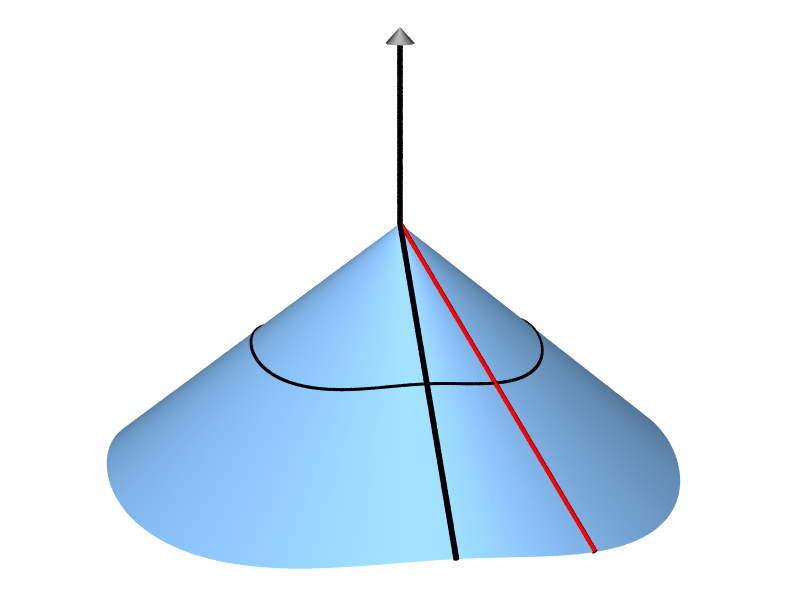}
	\put(100,40){$\Rightarrow$}
  \end{overpic}
  \begin{overpic}[width=3.75cm]{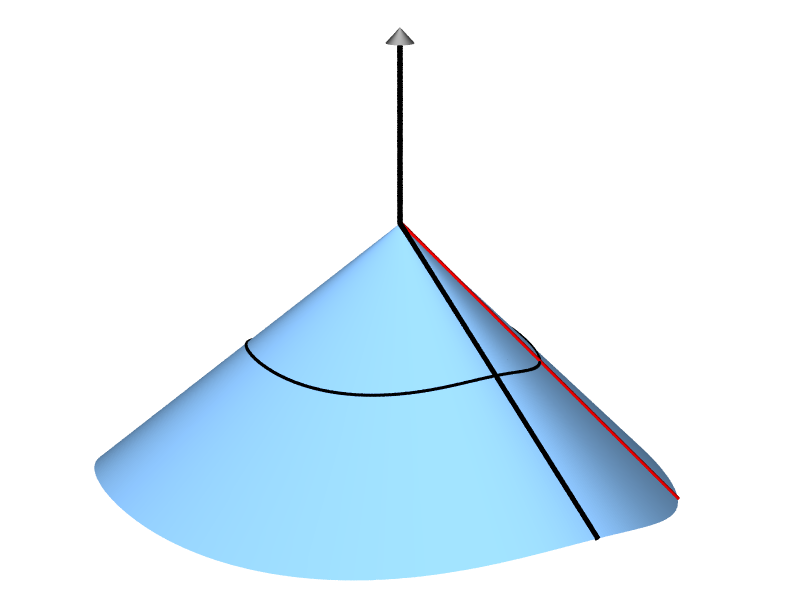}
  \end{overpic}
\caption{Snapshots of a flowing skirt for $\tau>0$. The cone and black line rotate around the vertical axis, while the material and red line additionally move with a zonal surface velocity along constant-$r$ curves.  The black and red lines coincide in the first snapshot.}
\label{fig:tau}
\end{figure}

In a flowing skirt, a traveling wave variable replaces the material zonal coordinate in the representation (\ref{cyl1}-\ref{cyl2}).  We consider $\bX(r, \eta, t) = r\uvc{u}(\eta, t)$, where $\eta \equiv s+\tau t$ and $\tau$ is a constant, signed frequency.  Thus, the material leads ($\tau > 0$) or lags ($\tau < 0$) the steady pattern, moving zonally along constant-$r$ curves.  At this point, we urge the reader who has not watched a video such as \cite{UrusoffRamosSkirtsYouTube} to do so now; we regret our inability to embed one in the text.  A few motion snapshots are shown in Figure \ref{fig:tau}.  When normalized by the frequency $\omega$, the magnitude of the traveling wave parameter $\tau$ is a Rossby number that indicates the relative importance of Coriolis terms in the equations of motion.  We will see that flowing skirts form a three-parameter family of shapes, of which the rigid skirts of Section \ref{rigidskirt} represent the ``geostrophic''\footnote{A confusing term for the present purposes.  We might suggest the more general ``somatostrophic'' to describe nearly rigid body motion.} limit.

The equivalent of \eqref{motion2}, again obtained from \eqref{motion}, is now
\begin{equation}\label{motiontravel}
\begin{split}
	&-\mu r^2 R \left(\omega + \tau\partial_\eta\Phi\right)^2	\left(\begin{array}{c}
												\cos [\omega t + \Phi]\\
												\sin [\omega t + \Phi]\\
												0
												\end{array}\right) 
	 +\mu r^2 \tau^2 	\left(\begin{array}{c}
					\partial_\eta^2R\cos [\omega t + \Phi]\\
					\partial_\eta^2R\sin [\omega t + \Phi]\\
					\partial_\eta^2Z
					\end{array}\right) \\
	&\quad +\mu r^2 \left[2\tau\partial_\eta R \left(\omega + \tau\partial_\eta\Phi\right)	+\tau^2R\partial_\eta^2\Phi\right]
	\left(\begin{array}{c}
	-\sin [\omega t + \Phi]\\
	\cos [\omega t + \Phi]\\
	0
	\end{array}\right) \\
	& \quad = \partial_r \left( r\sigma^{rr} \uvc{u}\right) + \partial_\eta \left( r^2 \sigma^{ss} \uvc{t} \right)  .
\end{split}
\end{equation}
The new terms scale with $r$ like the other acceleration terms, which is encouraging.  We note in passing that a gravitational term would scale differently.

Before proceeding, note that any shape corresponding to a static equilibrium of a generalized cone, that is, any solution to equation \eqref{motion2} with $\omega = 0$, now represents a family of solutions in the traveling wave variable, obtained by transforming the zonal stress along with the arc length variable: $\sigma^{ss} \rightarrow \sigma^{ss} + \mu\tau^2$, $s \rightarrow s + \tau t$.  The meridional stress $\sigma^{rr}$ is unaffected.  This is equivalent to what Routh observed in the context of inextensible strings \cite{Routh55}, namely that the centripetal forces induced by tangential motion along a curved string in an inertial frame are perfectly balanced by an increase in the string's tension. 
 In an inertial frame, adding a tangential velocity changes only the stress along the direction of motion, not the shape of the dynamical equilibrium solution.
However, in a rotating frame ($\omega \ne 0$), there will be an additional Coriolis term involving the product of $\omega$ and $\tau$, which does indeed generate new shapes.

Projection of \eqref{motiontravel} on $\partial_\eta\uvc{u}$ and integration in $\eta$ leads to the same expression \eqref{tproj} for $\sigma^{ss}$
\begin{equation}
	2\sigma^{ss} = -\mu\omega^2R^2 + C \, , \nonumber
\end{equation}
where we know that $C$ remains a constant as the $r$-scalings have not changed in the equation of motion.  The projection on $\uvc{u}$ followed by integration in $r$ gives the equivalent of \eqref{uproj},
\begin{equation}
	2r\sigma^{rr} = D(\eta) + 2 \int \left[\sigma^{ss} - \mu\left(\omega^2R^2 + \tau^2 + 2\omega\tau R^2\partial_\eta\Phi\right) \right]r^2 dr \, ,
\end{equation}
with $D(\eta)$ an arbitrary function of $\eta$.  The Coriolis forces act only along the normal and meridional directions.  We now project on the vector $\partial_t\uvc{u} - \tau \partial_\eta\uvc{u}$ and integrate over $\eta$ to find the new definition, replacing \eqref{Jz}, 
\begin{equation}
	2R^2\left[ \partial_\eta\Phi \left( \sigma^{ss} -\mu\tau^2\right) - \mu \omega \tau \right] \equiv J_z \, .
\end{equation}
In this expression, the Routhian symmetry under changes in centripetal, but not Coriolis, forces is immediately apparent.

Inserting for the stress and other quantities, normalizing as before, and defining the signed Rossby number $\hat{\tau} \equiv \frac{\tau}{\omega}$ yields the new equivalent of \eqref{zode},
\begin{equation}\label{zodetan}
	\left(\partial_\eta Z \right)^2 = 1 - Z^2 - \left[ \frac{\hat{J}_z + 2\hat{\tau}\left(1-Z^2\right)}{1 - Z^2 - \hat{C}+2\hat{\tau}^2} \right]^2 \, ,
\end{equation}
a system representing a non-degenerate three-parameter family of shapes, as a result of the Coriolis term linear in $\hat{\tau}$.  There are now tilted disk solutions corresponding to $\hat{J}_z = 2\hat{\tau}(2\hat{\tau}^2 - \hat{C})$, $4\hat{\tau}^2 < 1$, described by the circles $Z^2 + (\partial_\eta Z)^2=1 - 4\hat{\tau}^2$.  The presence and location of poles and outer wells are now governed by $\hat{C}- 2\hat{\tau}^2$ rather than $\hat{C}$. In contrast to the rigid skirt limit, the zonal stress in the central well may be tensile, restricted only by $\sigma^{ss} \le \mu\hat{\tau}^2$.  This should have important consequences for the stability of the solutions.  


The relevant cubic in $Z^2$ is now
\begin{equation}\label{cubictan}
	(1-Z^2)(1-Z^2-\hat{C}+2\hat{\tau}^2)^2 - \left[\hat{J}_z + 2\hat{\tau}\left(1-Z^2\right) \right]^2 =0 \, .
\end{equation}

The expression for the angle is
\begin{equation}\label{phitan}
	\Phi = \int^{\eta} \!\!\! d\tilde{\eta} \, \frac{\hat{J}_z+ 2\hat{\tau}\left[1-Z^2(\tilde{\eta})\right]}{[1-Z^2(\tilde{\eta})] [1-Z^2(\tilde{\eta})-\hat{C}+2\hat{\tau}^2]} \, ,
\end{equation}
and now $\Phi$ no longer need be monotonic in $s$, as $\hat{J}$ and $\hat{\tau}$ may differ in sign.  Indeed, they are always of opposite sign for multi-lobed deficit cones in the axisymmetric limit (see \ref{smallosc}).  Extending these limiting solutions to larger amplitude orbits, we encounter solutions that are both closed and self-intersecting.  Addition of a third parameter permits the appearance of additional extrema and inflection points in the potentials and orbits, reflected in qualitatively new types of solution beyond those displayed by rigid skirts.  An exploration of the full solution space will be discussed in a second paper.


\section{Example: a family of flowing deficit skirts}

\begin{figure}[here]
\subfigure{
	\begin{overpic}[width=3in]{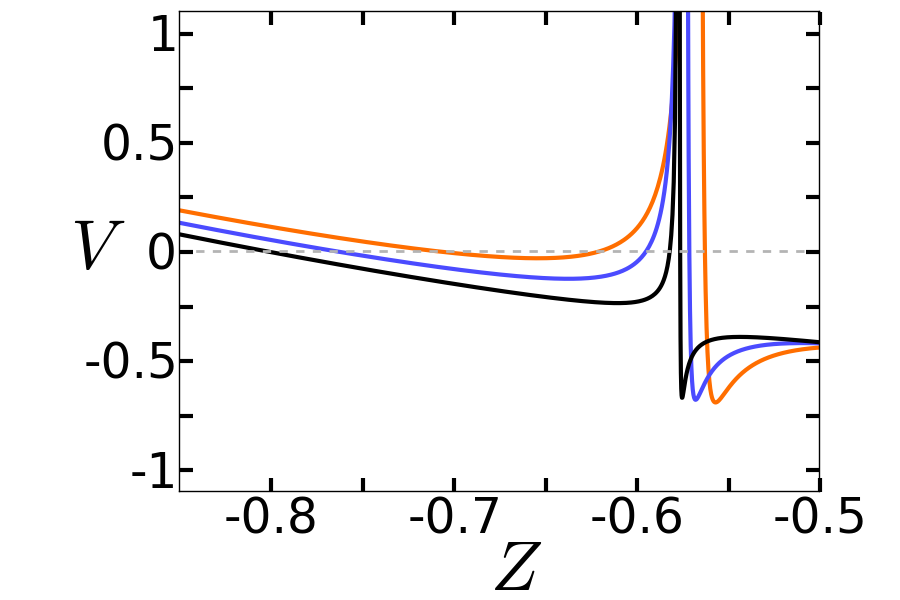}\label{top}
	\put(45,120){\Large{\subref{top}}}
	\end{overpic}
	}\hspace{.2in}
\addtocounter{subfigure}{1}
\raisebox{-.1in}{
\subfigure{
	\begin{overpic}[width=2.4in]{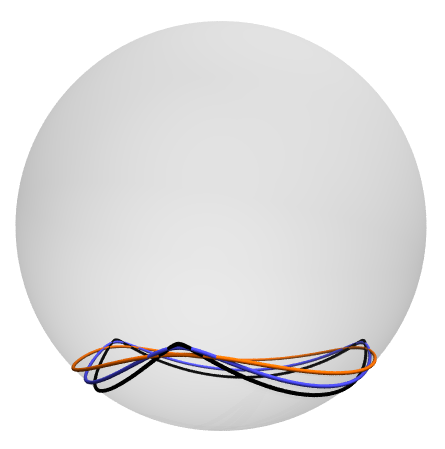}\label{threesols}
	\put(-22,128){\Large{\subref{threesols}}}
	\end{overpic}
	}
	}\\
\addtocounter{subfigure}{-2}
\subfigure{
	\begin{overpic}[width=3in]{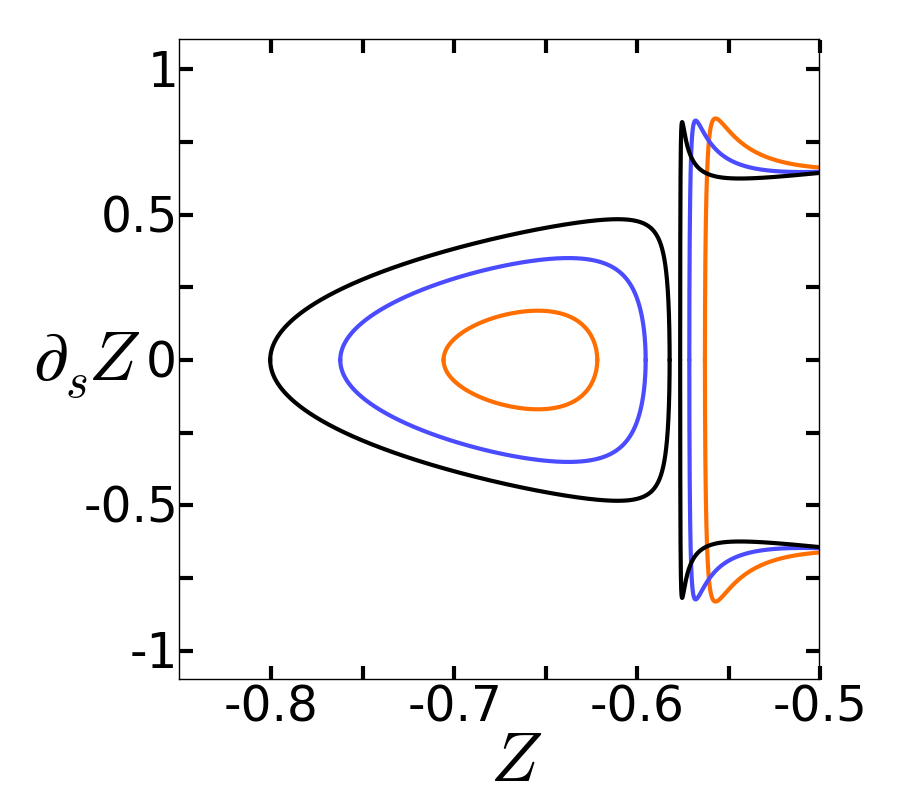}\label{bottom}
	\put(45,160){\Large{\subref{bottom}}}
	\end{overpic}
	}\hspace{-.1in}
\addtocounter{subfigure}{1}
\raisebox{0.15in}{
\subfigure{
	\begin{overpic}[width=1.5in]{dervishlingard.jpg}
	\end{overpic}
	}\hspace{-.1in}
\subfigure{
	\begin{overpic}[width=1.5in]{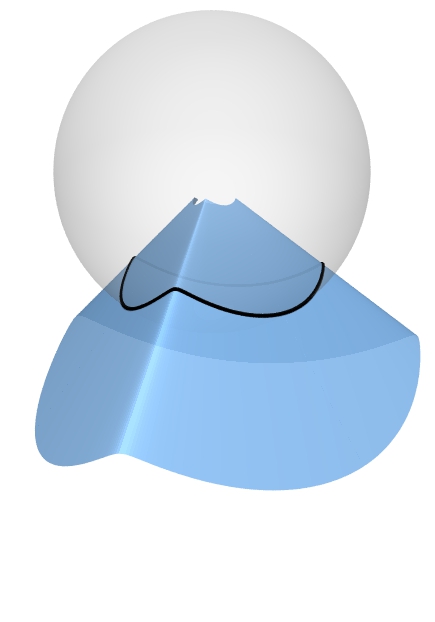}
	\end{overpic}
	}
}
\caption{A closeup of potentials \subref{top} and phase portraits \subref{bottom} showing outer-well outer-orbit solutions corresponding to three-lobed non-intersecting cones with a deficit angle of $\frac{\pi}{2}$.  Also visible is a portion of the central wells and orbits which correspond to the same trio of parameters $\hat{C}, \hat{J}_z, \hat{\tau}$ but not the conditions $\Delta = - \frac{\pi}{2}$ and $n=3$; the full system is symmetric about $Z=0$, so there is another outer solution with positive $Z$ not shown here.  In order of increasing sharpness and outer orbit area, $(\hat{C}, \hat{J}_z, \hat{\tau}) \approx {\color{Orange}(0.8991, -0.4585, 0.3325)} , {\color{blue}(0.8689, -0.4267, 0.3150)} , (0.8438, -0.3980, 0.2975)$.  The corresponding curves are drawn on the unit sphere \subref{threesols} and, below this, the surface corresponding to the most asymmetric of these is shown alongside a few dervishes. Photo adapted from the work of Matt Lingard (\texttt{www.flickr.com/photos/madrattling/2395433079/}), modified to blur faces.
}
\label{threelobe}
\end{figure}

A real cone-skirt is constructed from a given quantity of fabric, which quantity determines the strength of its singularity.  Here we display a family of solutions with fixed deficit angle $\Delta = - \frac{\pi}{2}$ and number of lobes $n=3$.  We restrict ourselves to those physically realistic solutions that do not self-intersect.  These solutions, three of which are shown in Figure \ref{threelobe}, occupy a relatively small region of parameter space which has a limiting axisymmetric solution of maximum positive $\hat{\tau}$ and $\hat{C}$ and minimum negative $\hat{J}_z$.  As the parameter values shift away from this limit, the system develops larger and more asymmetric orbits in phase space, corresponding to a greater asymmetry between the peaks and troughs of the skirt in real space.  The peaks become quite sharp and the troughs approach geodesic circular arcs, representing the planar faces of a three-sided pyramid. If one follows this family of solutions further beyond those shown, the outer orbits develop inflected ``ears'' like the central orbits, and the skirt peaks develop into self-intersecting loopty-loops.  
The physically realistic parameter space shrinks rapidly with increasing number of lobes.  All of these solutions correspond to backward-traveling waves in material coordinates, $\hat{\tau} > 0$, as in the perturbative analysis of \ref{smallosc}.
 
The resemblance of these solutions to real whirling skirts is better than one might expect, given that gravity has been neglected.  However, a detailed investigation of pattern selection and stability is warranted before saying more.

\section{Further discussion and summary}

There is a broader physical context in which to interpret our results: non-axisymmetric patterns are common on rotating solids, fluids, and interfaces.  Their existence may be variously ascribed, to propagating instabilities such as atmospheric Rossby waves on jet streams \cite{White90, PolvaniDritschel93}, spiral waves in shallow water \cite{Fridman85}, barotropic-inertial formation of secondary tornado vortices \cite{Snow78, StaleyGall79} and faceted cyclone eyewalls \cite{Schubert99}, or to the presence of exact dynamical equilibria such as local ``modon'' solutions of the barotropic vorticity equation \cite{Verkley84, White90} and global dihedrally symmetric forms including smoothed polygonal vortices \cite{Vatistas90, Bergmann11} and stellated bubbles \cite{WegmannCrowdy00} in inviscid fluids and trochoidal rosettes in (self-intersecting) strings \cite{Fusco84}.  All of these systems feature the freedom of the material to flow within the body independently of its rotation.  The admissible flows in unstretchable materials preserve local distances, so they are far more restrictive than those of, \emph{e.g.}, a spherical liquid layer subject to the Euler equations.  Indeed, for these solids, flows only exist on surfaces whose symmetries provide them with Killing fields.  Such surfaces, which include those with axisymmetric and developable metrics, share dynamical similarities with strings which we hope to explore in future work.

A shortcoming of the current formulation, with its normalization by the rotational frequency of the fixed shape, is its failure to clearly display the expected pairwise existence of solutions to what is essentially a wave equation.  This issue appears to be complicated further by the boundary condition implied by closure, which may select just one wave from a pair.  The asymmetric effects of a Coriolis term on wave pairs presents an interesting topic to examine in the general context of flexible body dynamics.  Another issue we should point out is that comparing energies and assessing the stability of shapes rotating at a particular frequency is unphysical, as the dancer or scientist will fix the \emph{material} rather than the shape frequency.  The question to ask is ``what shape and associated frequency will be selected for a given velocity imposed at the inner boundary?''

In conclusion, we have provided an analytical description of the steady configurations of a inextensible, flexible, \emph{flowing}, generalized-conical sheet rotating about a fixed axis, in the form of a three-parameter dynamical system describing the geometry.  This minimal model of metrically constrained dynamics displays the qualitative features of coherent patterns observed on a whirling skirt of fabric.  A quantitative study of selection, stability, and the choreography of bifurcations throughout the full solution space will be treated in subsequent publications and in conjunction with experiments.


\section*{Acknowledgments}

Initiation of this work was facilitated by the hospitality and support of the ICPM at Universit\'{e} de Lorraine.  JG acknowledges DGAPA PAPIIT grant IN114510-3 and CONACyT grant 180901.  JAH thanks C D Santangelo for support during a postdoctoral appointment through US NSF grants DMR 0846582 and DMR 0820506, the MRSEC on Polymers at the University of Massachusetts, Amherst.


\appendix
\section{Small oscillation analysis of solutions}\label{smallosc}

Here we treat the full three-parameter flowing skirt problem in the nearly axisymmetric limit.  The zonal evolution of $Z$ is taken to be a harmonic oscillation around a constant value $Z_0$, corresponding to a small circular orbit in phase space.  We must consider two different perturbation expansions, depending on whether or not we perturb around the plane $Z_0 = 0$.  
For the planar case, $\Phi \approx \eta$ and we consider the expansion
\begin{align}\begin{split}
	\hat{J}_z &\equiv J_0 + 0 + \epsilon^2 J_2 + \ldots \\
	\hat{C} &\equiv C_0 + 0 + \epsilon^2 C_2 + \ldots \\
	\hat{\tau} &\equiv T_0 + 0 + \epsilon^2 T_2 + \ldots \\
	Z &\equiv 0 + \epsilon A\cos q\eta + \epsilon^2 Z_2(\eta) + \ldots \, ,
\end{split}\end{align}
which leads immediately to $J_0 + 2T_0 = 1-C_0 +2T_0^2$.  Equation \eqref{zodetan} is trivially satisfied to first order in $\epsilon$.  At second order, eliminating trigonometric terms leads to $q^2 = 1 + \frac{2(1-2T_0)}{1-C_0+2T_0^2} = 1 + \frac{2(1-2T_0)}{J_0 + 2T_0}$ and some additional relationships between the amplitude $A$ and higher order terms. 

Closure without self-intersection requires integer values $n$ of $\frac{2\pi + \Delta}{2\pi}q \approx q$.  We ignore shifts ($n=0$) and tilts ($n=1$), examining only those solutions with two or more lobes.  In the rigid limit $\hat{\tau} = T_0 = 0$, $J_0 = 1 - C_0$ and there exists a discrete spectrum of solutions with 
$C_0 = \frac{q^2-3}{q^2-1} $.  Because $q \ge 2$, $C_0 < 1$ and the solutions live in an isolated central well with infinitely high barriers.  From each of these solutions emerges a continuous family for nonzero positive or negative values of $T_0$.

For expansions around cones $Z=Z_0$, where $Z_0$ is a nonzero root of \eqref{cubictan}, $\Phi \approx \frac{\eta}{\sqrt{1-Z_0^2}}$ and we write
\begin{align}\begin{split}
	\hat{J}_z &\equiv J_0 + 0 + \epsilon^2 J_2 + \ldots \\
	\hat{C} &\equiv C_0 + 0 + \epsilon^2 C_2 + \ldots \\
	\hat{\tau} &\equiv T_0 + 0 + \epsilon^2 T_2 + \ldots \\
	Z &\equiv Z_0 + \epsilon A\cos q\eta + \epsilon^2 Z_2(\eta) + \ldots \, ,
\end{split}\end{align}
which leads to $(1-Z_0^2)(1-Z_0^2-C_0+2T_0^2)^2 - [J_0 + 2T_0 (1-Z_0^2) ]^2 =0$.  The first order condition from \eqref{zodetan} is now the less-restrictive $J_0^2 + 2(1-Z_0^2-C_0)(1-Z_0^2)^2 = 0$.  At second order, we now have $q^2 = \frac{4Z_0^2[2C_0-3(1-Z_0)^2]}{(1-Z_0^2-C_0+2T_0^2)^2} = \frac{4Z_0^2[J_0^2-(1-Z_0^2)^3]}{(1-Z_0^2) [J_0 + 2T_0 (1-Z_0^2) ]^2}$.  This implies the restrictions $C_0 \ge \frac{3}{2}(1-Z_0^2) > 0$ and $J_0^2 \ge (1-Z_0^2)^3$, with equality corresponding to a shifted solution that we ignore.

Because $\left(\frac{2\pi + \Delta}{2\pi}\right)^2 \approx 1 - Z_0^2$, we now seek $(1-Z_0^2)q^2 = n^2$ with integer-valued  $n$.  Hence
\begin{equation}
	n^2 = \frac{4Z_0^2(1-Z_0^2)\left[2C_0-3(1-Z_0^2)\right]}{(1-Z_0^2-C_0+2T_0^2)^2} =  \frac{4Z_0^2\left[J_0^2-(1-Z_0^2)^3\right]}{ \left[J_0 + 2T_0 (1-Z_0^2) \right]^2} \, .
\end{equation} 
Let $J_0 \equiv u(1-Z_0^2)^\frac{3}{2}$, $C_0 \equiv v(1-Z_0^2)$, and $T_0 \equiv w(1-Z_0^2)^\frac{1}{2}$, with $u^2 \ge 1$ and $v \ge \frac{3}{2}$.  Then the zeroth, first, and second order conditions are
\begin{align}
	(1 - v + 2w^2)^2 - (u+2w)^2 &= 0 \, , \\
	u^2 + 2(1-v) &= 0 \, , \\
	n^2 &= 4Z_0^2 \frac{u^2-1}{(u + 2w)^2}  \, ,
\end{align}
which system has solutions
\begin{align}
	u &= -\frac{n^2+Z_0^2}{n^2-Z_0^2} \, , \\
	v &=1 + \frac{(n^2+Z_0^2)^2}{2(n^2-Z_0^2)^2} \, , \\
	w&= \frac{n^2 - 3Z_0^2}{2(n^2-Z_0^2)} \, . \label{weq}
\end{align}
The signs of $u$ and $w$ may be simultaneously switched, but this simply inverts the sign of $\Phi$ and does not provide a new solution.  Again we consider $n \ge 2$.  The closure condition appears to have chosen only the backward-traveling wave ($w>0$) solution of the wave equation of motion.  In the rigid limit, $w = 0$ and because $Z_0^2 \le 1$, equation \eqref{weq} cannot be satisfied.  Thus, unlike the $Z_0 = 0$ planar case, we find no rigidly rotating perturbative solutions with multiple lobes.  For large values of $n$, all of the solutions approach $u \rightarrow -1, v \rightarrow \frac{3}{2}, w \rightarrow \frac{1}{2}$.  If $C_0 - 2T_0^2 = (1+4Z_0^2)(1-Z_0^2) \le 1$, that is, if $|Z_0| \ge \frac{1}{2}\sqrt{\frac{3}{2}}$, the corresponding wells will have infinitely high barriers.


\section*{References}

\bibliographystyle{unsrt}

\end{document}